\documentclass{ws-p9-75x6-50}

\newcommand{\dsla}{\partial\hspace{-1.5mm}/}
\newcommand{\psla}{p\hspace{-1.5mm}/}

\newcommand{\Asla}{A\hspace{-2.0mm}/}

\newcommand{\nn}{\nonumber}

\begin{document}

\title{
ERG and Schwinger-Dyson Equations\\
-- Comparison in formulations and applications --
}

\author{Haruhiko Terao}

\address{
Institute for Theoretical Physics, Kanazawa University, \\
Kanazawa 920-1192, Japan 
\\E-mail: terao@hep.s.kanazawa-u.ac.jp}

\maketitle

\abstracts{
The advantageous points of ERG in applications to non-perturbative
analyses of quantum field theories are discussed 
in comparison with the Schwinger-Dyson equations. 
First we consider the relation between these two formulations 
specially by examining the large N field theories.
In the second part we study the phase structure of 
dynamical symmetry breaking in three dimensional QED 
as a typical example of the practical applications.
}
\section{Introduction}
It has been a desire to have practically useful frameworks for
analytical studies of non-perturbative dynamics of field theories.
So far the Schwinger-Dyson (SD) equations have been mostly
used in many aspects (except for some supersymmetric theories).
The gap equations and also the loop equations are known as the typical
examples of them.

However Exact Renormalization Group (ERG)\cite{ERG,Polchinski,Legendre} 
is expected to offer
a equally or more powerful non-perturbative method.
These two formulations share the following commomn features.
The correlation functions are given as those solutions.
Regularization is necessary to define the equations.
The full equations are given in functional forms,
therefore approximations are inevitable in practical calculations.

On the other hand the characteristics of the ERG are as follows.
ERG gives the RG flows for the effective couplings, while
the SD equations give the order parameters in terms of 
the bare couplings.
Therefore the phase diagrams are given in the effective 
(renormalized) coupling space in the ERG formalism.
Also the fixed points, the critical exponents
and the renormalized trajectories (continuum limit) are directly
evaluated.
It should be also noted that the ERG allows systematic 
improvement of approximations: the derivative expansion,
while the systematic treatment has not been known in general
for the SD equations. 

In this paper we are going to see the interrelations 
between these formulations and compare the characteristics 
more closely. In the first part we discuss the relations in
formulations by considering large N field theories.
In the second part the dynamical symmtry breaking in QED$_3$
is exmained for the comparison in the practical applications.  

\section{SD and ERG in Large N Field Theories}
\subsection{The large N vector model}
Let us begin with the large N vector models. It is well known
that this class of models may be solved by using the so-called
auxiliary field method, or the Hubbard-Stratonovich transformation.
Here, however, we restrict ourselves to the ERG and SD approaches.
The euclidean bare action is given by
\be
S_b = \int d^d x~
\frac{1}{2}(\partial_{\mu}\phi)^2 +\frac{1}{2}m^2\phi^2
+\frac{g}{8N}(\phi^2)^2, 
\label{eq.1}
\ee
where $\phi^2=\phi^a\phi^a~~~(a=1,\cdots,N)$.
The cutoff effective action $\Gamma[\phi]$ satisfies
the so-called Legendre flow equation\cite{Legendre}
given (apart from the canonical scaling part) by
\be
\frac{\partial \Gamma_{\Lambda}[\phi]}{\partial \Lambda}
~=~ \frac{1}{2}\mbox{Tr}
\left(\frac{\partial \Delta^{-1}}{\partial \Lambda}
\left[\Delta^{-1} \delta_{ab} + 
\frac{\delta^2 \Gamma_{\Lambda}[\phi]}{\delta \phi^a \delta \phi^b}
\right]^{-1}
\right),
\label{eq.2}
\ee
where $\Delta(p)$ is the cutoff propagator.

In the large N limit we obtain the ERG equation
\be
\frac{\partial \Gamma_{\Lambda}[\rho]}{\partial \Lambda}
~=~ \frac{1}{2} \mbox{Tr}
\left(\frac{\partial \Delta^{-1}}{\partial \Lambda}
\left[\Delta^{-1} + 
\frac{\delta \Gamma_{\Lambda}[\phi]}{\delta \rho }
\right]^{-1}
\right)
\label{eq.3}
\ee
by redefining $\Gamma_{\Lambda} \rightarrow N \Gamma_{\Lambda}$ and 
$\rho = \phi^2/2N$.
It is seen that the flow equation for the potential is 
exactly extracted in this case as\cite{largeN}
\be
\frac{\partial V(\rho)}{\partial \Lambda}
~=~ \frac{1}{2}\int \frac{d^dp}{(2\pi)^d} 
\frac{\partial \Delta^{-1}(p)}{\partial \Lambda}
\frac{1}{\Delta^{-1}(p) + V'(\rho)}.
\label{eq.4}
\ee
Below we see explicitly that the solution of the ERG equation indeed
satisfies the SD equation also.

The SD equations are nothing but the identities
between correlation functions followed by the trivial equation
\be
0=
\int{\cal D} \phi 
\frac{\delta}{\delta \phi^a(x)}
\left( O[\phi]e^{-S[\phi]} \right)
= 
\int{\cal D} \phi 
\left( \frac{\delta O[\phi]}{\delta \phi^a(x)}
- \frac{\delta S[\phi]}{\delta \phi^a(x)}O[\phi]
\right)e^{-S[\phi]}.
\label{eq.5}
\ee
However it is necesarry to perform regularization in order
to make the equations meaningful. Here we introduce 
cutoff to the propagator just as done for the Legendre flow
equations:
\be
S[\phi] = 
-\frac{1}{2}\phi^a\cdot\Delta^{-1}\cdot\phi^a -S_b^{\rm int}[\phi]
\label{eq.6}
\ee
If we set $O = \phi^b(y)$ for example, then we obtain 
\be
(\Delta^{-1} + m^2)\langle \phi^a(x)\phi^b(y)\rangle
+\frac{g}{2N}
\langle \phi^2(x)\phi^a(x) \phi^b(y)\rangle
=\delta^{ab} \delta^{(d)}(x-y).
\label{eq.7}
\ee
These equations are not closed, therefore we need to handle the
infinite set of the SD equations in general.
The factorization property in the large N limit reduces 
the above SD equation to the one for the two point functions:
\be
\left(
\Delta^{-1} + m^2 + g\rho + \frac{g}{2N}\langle \phi^2(x)\rangle_c
\right)
\langle \phi^a(x)\phi^b(y)\rangle = \delta^{ab} \delta^{(d)}(x-y),
\label{eq.8}
\ee
where the subscript c indicates the connected part. 
It is seen from this equation that the mass function defined by
$
\Sigma = m^2 + g\rho + (g/2N)\langle \phi^2(x)\rangle_c
$
satisfies the so-called gap equation
\be
\Sigma = m^2 + g\rho + \frac{g}{2} \int \frac{d^dp}{(2\pi)^d} 
\frac{1}{\Delta^{-1} + \Sigma}.
\label{eq.9}
\ee
Here it should be noted that the mass function is related to the
cutoff effective potential as
\be 
\Sigma(\rho) = \frac{\partial^2 V(\phi)}{\partial \phi^2} 
= \frac{\partial V(\rho)}{\partial \rho}.
\label{eq.10}
\ee

By solving the gap equation the mass function is given in terms
of $\rho$ and the cutoff scale $\Lambda$.
The derivatives $\partial \Sigma / \partial \rho$
and $\partial \Sigma / \partial \Lambda$ satisfies the
following equations,
\bea
\frac{\partial \Sigma}{\partial \rho} &=&
g - \frac{g}{2} \int \frac{d^dp}{(2\pi)^d} 
\frac{1}{(\Delta^{-1} + \Sigma)^2}
\frac{\partial \Sigma}{\partial \rho},
\label{eq.11}\\
\frac{\partial \Sigma}{\partial \Lambda} &=&
- \frac{g}{2} \int \frac{d^dp}{(2\pi)^d} 
\frac{1}{(\Delta^{-1} + \Sigma)^2}
\left(
\frac{\partial \Delta^{-1}}{\partial \Lambda}
+ \frac{\partial \Sigma}{\partial \Lambda}
\right).
\label{eq.12}
\eea
Further we may rewritten the scale dependence of the mass function
as
\be
\frac{\partial \Sigma}{\partial \Lambda} =
- \frac{g}{2} \int \frac{d^dp}{(2\pi)^d} 
\frac{\partial \Delta^{-1}}{\partial \Lambda}
\frac{1}{(\Delta^{-1} + \Sigma)^2}
\frac{\partial \Sigma}{\partial \rho},
\label{eq.13}
\ee
by using Eq.~(\ref{eq.9}) and (\ref{eq.11}). 
This differential equation is
found to be just equivalnt to the ERG equation for the effective
potential (\ref{eq.4}). 
Thus the equaivalence in the large N limit is shown.

\subsection{The Gross-Neveu model}
In this case as well the equaivalence of
ERG and SD may be shown by repeating the similar argument
given above.
However we should pay 
attention to that the effective potential treated by the
ERG formulation is not the potential in terms of the 
order parameter given by the fermion composite. 
Let us mention the ERG analysis briefly also for the later 
conveniences.

We start with the bare action given by
\be
S_b = \int d^d x~
\bar{\psi}_i \Delta^{-1} \psi^i - \frac{G}{2N}
(\bar{\psi}_i\psi^i)^2,~~~~~(i=1,\cdots,N),
\label{eq.14}
\ee
where we introduced the cutoff propagator
$\Delta(p)=C(p^2/\Lambda^2)/i\psla$.
Note that chiral symmetry is preserved by this regularization. 
In the large N limit the ERG equation for the cutoff effective
potential $V(\psi, \bar{\psi}; \Lambda)$ may be exactly
derived again.
By redefining $V \rightarrow NV$ it is found to be
\be
\frac{\partial V(\sigma)}{\partial \Lambda}
= \int \frac{d^dp}{(2\pi)^d} \mbox{Tr}
\left(\frac{\partial \Delta^{-1}}{\partial \Lambda}
\frac{1}{\Delta^{-1} + V'(\sigma)}
\right),
\label{eq.15}
\ee
where $\sigma$ denotes a product of the classical fields, 
$\bar{\psi}\psi$.
It should not be confused with $\langle \bar{\psi}\psi \rangle$.

Now our interest is to see dynamical mass generation in this model.
In the SD approach the mass function $\Sigma$ defined by
\be
\langle \psi_{\alpha}^i(p)\bar{\psi}_{j \beta}(-p)\rangle
=\delta_{\alpha\beta}\delta_j^i \frac{1}{\Delta^{-1}(p) + \Sigma(p)},
\label{eq.16}
\ee
is examined.
Here it should be noted that we first assume the order parameter
apriori to see the critical phenomena in the SD approach.
In the large N limit the gap equation is found to be
\be
\Sigma = -G\sigma - G \int \frac{d^dp}{(2\pi)^d}\mbox{Tr}
\frac{1}{\Delta^{-1}+\Sigma},
\label{eq.17}
\ee
where we kept the classical fields $\sigma$.
The critical (bare) coupling for the dynamical chiral 
symmetry breaking is found by 
(non-)exsistence of non-trivial solutions for this gap equation. 

We may derive the scale dependence of the mass function by
considering $\partial \Sigma/\partial \sigma$ and
$\partial \Sigma/\partial \Lambda$ to the gap equation.
The resultant ERG for the mass function turns out to be
\be
\frac{\partial \Sigma}{\partial \Lambda} =
- \int \frac{d^dp}{(2\pi)^d}\mbox{Tr}
\left( \frac{\partial \Delta^{-1}}{\partial \Lambda}
\frac{1}{(\Delta^{-1} + \Sigma)^2 }
\right)
\frac{\partial \Sigma}{\partial \sigma}.
\label{eq.18}
\ee
Once we note that the mass function is related with
the effective potential as $\Sigma(\sigma) = dV/d\sigma$,
then the equaivalence of these formulations is readily seen.
\footnote{
Actually the gap equation leads to also the solutions
corresponding to unstable vacua. Further study is needed to
understand these solutions in the ERG point of view.}

However the ERG has a great advantage to find the phase structures
and also the critical exponents compared with the SD approaches. 
If we perform the operator expansion of the effective potential
into
\be
V(\sigma; \Lambda)=-\frac{1}{2\Lambda^{d-2}} G(\Lambda)\sigma^2
+ \frac{1}{8\Lambda^{3d-4}}G_8(\Lambda) \sigma^4 + \cdots,
\label{eq.19}
\ee
then the beta function for each coupling is derived
by substituting into Eq.~(\ref{eq.15}).
It is found that the effective 4-fermi coupling $G$ is
subject to the ERG equation isolated from other couplings:
\be
\beta_{G}=\Lambda\frac{dG}{d\Lambda}
= (d-2)G - A G^2,
\label{eq.20}
\ee
where $A$ is a  cutoff scheme dependent constant.
This beta function has two fixed points:
$G^*=0$ (IR attractive) and $G^*=(d-2)/A$ (IR repulsive).
The IR repulsive fixed point gives the critical coupling
of the chiral symmetry breaking and there are found to be
broken phase and unbroken one. It is also quite easy
to see the anomalous dimensions of the operators
$\bar{\psi}\psi$, $(\bar{\psi}\psi)^2$ and so on\cite{ours1}.

However it may seem curious how the dynamical mass can
be generated in the broken phase in the ERG formulation.
Indeed the chiral symmetry prohibits the mass term 
and any symmetry breaking operators to appear
in the effective potential $V(\sigma,\Lambda)$.
It is found that the operator expansion given by Eq.~(\ref{eq.19})
is not proper in order to see that composite order parameter
\cite{ours2}.

The dynimacal mass is rather evaluated as the mass function
at $\sigma=0$:
\be
m_{\rm eff} 
= \lim_{\Lambda \rightarrow 0} \Sigma(\sigma,\Lambda) |_{\sigma=0}
= \lim_{\Lambda \rightarrow 0} V'(\sigma,\Lambda) |_{\sigma=0}.
\label{eq.21}
\ee
If we solve the ERG equation for the effective potential
$V(\sigma;\Lambda)$, then it is found that
the potential is evoluted to be non-analytic at the origine 
due to the IR singularity of massless fermion loops\cite{ours2}.
On the other hand the fermion condensate
$\langle \bar{\psi}\psi\rangle$ is also obtained by introducing
the bare mass $m_0$ in the original action.
Since $m_0$ plays a role of source for the fermion composite,
$\langle \bar{\psi}\psi\rangle $ is evaluated by
\be
\langle \bar{\psi}\psi\rangle 
= \lim_{\Lambda \rightarrow 0}
\left.
\frac{\partial V(\sigma, m_0, \Lambda)}{\partial m_0}
\right|_{\sigma=0+}.
\label{eq.22}
\ee
Thus we need to analyze the whole potential, 
not only the four fermi coupling, to see generation of 
the order parameters of the dynamical symmetry breaking. 
However the phase structure of dynamical symmetry breaking
is immediately found out from the RG flows of couplings.
\footnote{
It has been found that if we introduce the composite
operators corresponding to the fermion condensate by
extending the theory space, then the operator expansion scheme
works quite well\cite{ours2,Ellwanger1}.
}

\subsection{Formal equivalence}
It is also shown formally that 
the solutions of the ERG equation neccesarily satisfy the 
SD equation in generic field theories.
\footnote{
The author became aware after the conference that the same argument
has been already given by Ellwanger {\it et al}\cite{Ellwanger2}.
}
First we define the cutoff generating functional
\be
Z_{\Lambda}[J]
= e^{W_{\Lambda}[J]}
= \int {\cal D}\phi e^{-S_{\Lambda} + J\cdot\phi}
\label{eq.23}
\ee
by using the regularized action
\be
S_{\Lambda}[\phi] = 
\int \int d^dx d^d y~
\frac{1}{2}\phi(x) \Delta(x-y;\Lambda) \phi(y) + S[\phi].
\label{eq.24}
\ee
The variation of the generating functional under shift of 
the cutoff $\Lambda$ is given as
\be
\frac{\partial}{\partial \Lambda}Z_{\Lambda}[J]
=\frac{1}{2}\int \int d^dx d^d y~
\frac{\delta}{\delta J(x)} 
\frac{\partial \Delta}{\partial \Lambda} 
\frac{\delta}{\delta J(y)} 
Z_{\Lambda}[J],
\label{eq.25}
\ee
which may be rewritten also into the Polchinski equation\cite{Polchinski}.
On the other hand the general SD equations are represented 
in terms of the source function as
\be
0= 
\left(J(x) - \frac{\delta S_{\Lambda}}{\delta \phi(x)}
\left[\frac{\delta}{\delta J}\right]
\right)Z_{\Lambda}[J].
\label{eq.26}
\ee

Now suppose that the generating functional $Z_{\Lambda}[J]$ 
satisfies the SD equation derived for the action $S_{\Lambda}$.
Under variation of the scale
$\Lambda \rightarrow \Lambda + \delta\Lambda$, the generating
functional is shifted by
\be
Z_{\Lambda+\delta\Lambda}[J] =
Z_{\Lambda}[J] + \frac{1}{2}\int \int d^dx d^d y~
\frac{\delta}{\delta J(x)} 
\frac{\partial \Delta}{\partial \Lambda} 
\frac{\delta}{\delta J(y)} 
Z_{\Lambda}[J]\delta \Lambda.
\label{eq.27}
\ee
Then we may show that the generating functional at scale
$\Lambda + \delta\Lambda$ indeed satisfies the SD equation
deduced from the action $S_{\Lambda+\delta\Lambda}$:
\be
\left(J(x) - \frac{\delta S_{\Lambda+\delta\Lambda}}{\delta \phi(x)}
\left[\frac{\delta}{\delta J}\right]
\right)Z_{\Lambda+\delta\Lambda}[J]=0,
\label{eq.28}
\ee
by noting thet $Z_{\Lambda}[J]$ satisfies Eq.~(\ref{eq.26}).
If we perform both UV and IR cutoff to define the generating functional,
then the solutions of the SD equation are obtained by removing the IR  
cutoff. Therefore it is seen generally that the solutions of the 
ERG equation at the IR limit should satisfies the SD equation.

\subsection{The large N matrix model}
In this subsection we consider to apply the ERG method to the 
large N matrix model\cite{matrix,Onoda} given by the action
\be
S= N
\int d^dx \frac{1}{2}\mbox{Tr}[(\partial_{\mu}\phi)^2]
+\frac{1}{2}m^2 \mbox{Tr}[\phi^2]
+\frac{g}{4}\mbox{Tr}[\phi^4],
\label{eq.29}
\ee
where $\phi$ is a N $\times$ N hermitian matrix.
\footnote{
In the large N limit the multi trace opertaors are not generated 
from the single trace ones through radiative corrections.
Therefore let us strat with this type of actions. 
General cases containing the multi trace operators may be
analyzed similarly\cite{Onoda}.  
}
The SD equations for the matrix model are known as the loop
equations. For example we obtain in the large N limit
\be
 (-\partial_x + m^2)\langle \mbox{Tr}\phi(x) \phi(y)\rangle
+g \langle \mbox{Tr}\phi^3(x) \phi(y)\rangle
=\frac{1}{N} \sum_{p=0}^{n-1}
\langle \mbox{Tr}\phi^p(x) \rangle
\langle \mbox{Tr}\phi^{n-p-1}(x)\rangle.
\label{eq.30}
\ee
However it has not been known how to treat such equations.
\footnote{
Except for the 0 dimensional model.
}

Let us consider the ERG in the LPA approximation\cite{LPA}, 
in which radiative corrections including the derivatives are discarded.
The generic Legendre flow equation leads to the ERG for
the effective potential $V(\phi)$ as
\be
\frac{\partial V}{\partial \Lambda} =
\frac{1}{2} \int \frac{d^dp}{(2\pi)^d}
\frac{\partial \Delta^{-1}}{\partial \Lambda}
\mbox{Tr}
\left[\Delta^{-1}\delta_{il}\delta_{jk}
+\frac{\delta^2 V}{\delta \phi_{ij}\delta\phi_{kl}}
\right]^{-1},
\label{eq.31}
\ee
apart from the canonical scaling terms.
As is seen later on the corrections are limited to the single
trace operators in the large N limit\cite{matrix,Onoda,gauge}. 
Therefore we may
suppose that the potential is consist of single trace oprators:
\be
V(\phi) = \sum_{n=1}^{\infty} a_n(\Lambda) \mbox{Tr} \phi^{2n}.
\label{eq.32}
\ee
Then in the second derivatives in Eq.~(\ref{eq.31}),
\be
\frac{\delta^2 V}{\delta \phi_{ij}\delta\phi_{kl}}
=\sum_{n=1}^{\infty} 2n a_n(\Lambda) 
\left\{
\delta_{li}(\phi^{2n-2})_{jk}
+(\phi^{2n-2})_{li}\delta_{jk}
+\sum_{p=1}^{2n-3}(\phi^{p})_{li}(\phi^{2n-p-2})_{jk}
\right\},
\label{eq.33}
\ee
only the first and the second terms can contributes to the
radiative correcitons in the large N limit.
Therefore the trace in the ERG may be evaluated as
\be
\mbox{Tr}
\left[\Delta^{-1}+\frac{\delta^2 V}{\delta \phi \delta\phi}\right]^{-1}
= 2N \mbox{Tr}\left(\left[
\Delta^{-1} + \sum_{n=1}^{\infty} 2n a_n \phi^{2n-2} 
\right]^{-1}\right).
\label{eq.34}
\ee
Thus only the single trace operators are found to appear through 
the corrections.
In Fig.~1 the large N leading corrections are shown schematically.
It is realized also that solving the ERG equation generates
planar diagrams. 
\begin{figure}[h]
\begin{center}
\epsfxsize=0.4\textwidth
\leavevmode
\epsffile{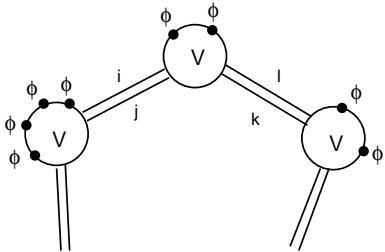}
\caption{The large N leading corrections are shown schematically.}
\end{center}
\end{figure}

We may deduce the beta functions for the couplings $a_n$ as follows.
If we define a function depending on $a_n$ as
\be
v(y) = \sum_{n=1}^{\infty} a_n y^{n},
\label{eq.35}
\ee
then the beta functions are obtained by expanding the equation,
\be
\Lambda\frac{\partial v}{\partial \Lambda} =
N \int \frac{d^dp}{(2\pi)^d}
\Lambda \frac{\partial \Delta^{-1}}{\partial \Lambda}
\left[\Delta^{-1} + 2v'(y))\right]^{-1},
\label{eq.36}
\ee
which is found to be identical to the flow equation for the effective 
potential of the large N vector model. 
Therefore there are found a critical surface dividing two phases. 
Contrary to the vector model, however,
the LPA is not exact, since 
there are additional corrections from the derivative interactions. 
Namely the large N matrix model and the vector model belong to
different universality classes. 
Here we would like to remark that ERG is able to handle the matrix
models as well by applying the derivative expansion\cite{Onoda}.

\section{Dynamical Symmetry Breaking in QED$_3$}
Now we discuss the feature of ERG in applications to the
dynamical symmetry beraking by gauge interactions. 
Here we consider specially the application to QED$_3$, 
which serves a typical example\cite{Kubota}.

Dynamical symmetry breaking by gauge interaction is one of 
the most interesting non-perturbative phenomena not only 
in condensed matter physics but also in  particle physics. 
In practice many works using the SD equations has been done
mostly in the applications to such problems, for example,
chiral symmetry breaking in 4 dimensional gauge theories,
color superconductivity in high density QCD and so on.

In the SD approach we have to assume the order parameter
essential for the dynaminal symmetry breaking. For the chiral
symmetry breaking we examine the equation for the mass function.
In general, however, we do not know apriori which symmetry 
should be broken dynamically. This problem becomes important
in the case that there are more than two phases.
Indeed we will face with such a situation in considering
QED$_3$.
It should be mentioned also that the ladder approximation
scheme, which has been frequently used in the SD approach,
sffers from large gauge parameter dependence. Moreover, 
if we try to improve the approximation so as to
obtain gauge independent results, 
we are obliged to treat much more complicated equations. 
In this section we examine QED$_3$ by the ERG equations
in comparing with such features in the SD approach.

Let us consider QED$_3$ with $N$ flavors of 4 component 
spinors $\psi^i$ ($i=1, \cdots, N$) without the Chern-Simons term.
The bare lagrangian is given by
\be
S_b= \sum_{i=1}^{N} \bar{\psi}_i(\dsla + e \Asla)\psi^i
+ \frac{1}{4} F_{\mu \nu}^2 -
  \frac{1}{2\xi}(\partial_{\mu}A_{\mu})^2.
\label{eq.37}
\ee 
Here we use the 4 by 4 $\gamma$ matrices given by 
\be
\gamma^0 = 
\left( 
\begin{array}{cc}
\sigma_3 & 0 \\
0 & -\sigma_3
\end{array}
\right),~
\gamma^1 = 
\left( 
\begin{array}{cc}
\sigma_1 & 0 \\
0 & -\sigma_1
\end{array}
\right),~
\gamma^2 = 
\left( 
\begin{array}{cc}
\sigma_2 & 0 \\
0 & -\sigma_2
\end{array}
\right),~
\gamma^3 = 
\left( 
\begin{array}{cc}
0 & -i \\
i & 0 
\end{array}
\right).
\label{eq.38}
\ee
We also introduce
$\gamma^5 =\gamma^0  \gamma^1  \gamma^2  \gamma^3$
and
$\tau= -i \gamma^5  \gamma^3$.
This action is invariant under the global U(2N) and also parity 
symmetry, which is made transparent by reformulating in terms of 2n 
2-component spinors $\chi_I$, ($I=1, \cdots, 2N$):
\be
\psi^i = 
\left(
\begin{array}{c}
\chi^i \\
\chi^{i+N}
\end{array}
\right),~~~~
\bar{\psi}_i = 
(\chi_i^{\dagger}\sigma_3, -\chi_{i+N}^{\dagger}\sigma_3)
=(\bar{\chi}_i, \bar{\chi}_{i+N})\tau.
\label{eq.39}
\ee
The 2-component field are transformed by the U(2N) matrix $U$ as
$\chi^I \rightarrow \chi'^I= U_J^I \chi^J $.
The parity transformation is defined by
$\psi \rightarrow \psi'=i\gamma^3\gamma^5\psi^i$.
Therefore 
$\bar{\psi}_i \gamma^{\mu} \psi^i =
\bar{\chi}_I \gamma^{\mu}\chi^I$
is invariant under the both symmetry.
The ordinary mass opertaor,
$\bar{\psi}_i \psi^i = 
\bar{\chi}_i \chi^i - \bar{\chi}_{i+N}\chi^{i+N}$
is parity even but not invariant under U(2N) transformation.
If this opertaor acquires a non-vanishing vaccum expectation value,
then U(2N) is spontaneously broken to U(N)$\times$U(N).
Thus we may regards this U(2N) symmetry as a sort of
chiral transformation.
While we find a U(2N) invariant operator,
$\bar{\psi}_i \tau \psi^i = 
 \bar{\chi}_I \chi^I $,
which is parity odd in turn.
Therefore non-vanishing expectation value of this operator
leads to spontaneous breakdown of the parity symmetry.
However it is expected from Vafa-Witten's theorem\cite{VafaWitten}
that parity is never broken in QED$_3$.

In section~1 we saw that the RG flows of the effective four fermi 
interactions are important to distinguish the phases.
All the local four-fermi operators invariant under 
U(2N) and parity tarnsformations are listed up as follows:
\bea
{\cal O}_P &=&
(\bar{\psi}_i\tau \psi^i)^2 = (\bar{\chi}_I\chi^I)^2 \nn \\
{\cal O}_V &=&
(\bar{\psi}_i\gamma^{\mu} \psi^i)^2 
= (\bar{\chi}_I\gamma^{\mu}\chi^I)^2 \nn \\
{\cal O}_S &=&
\bar{\psi}_i \psi^j\bar{\psi}_j \psi^i 
- \bar{\psi}_i \gamma^3 \psi^j \bar{\psi}_j \gamma^3\psi^i  
- \bar{\psi}_i \gamma^5 \psi^j \bar{\psi}_j \gamma^5\psi^i 
+ \bar{\psi}_i \tau \psi^j \bar{\psi}_j \tau \psi^i \nn \\
 &=& 2 \bar{\chi}_I\chi^J \bar{\chi}_J\chi^I  \nn \\
{\cal O}_{V'} &=&
\bar{\psi}_i \gamma^{\mu}\psi^j\bar{\psi}_j \gamma^{\mu}\psi^i 
- 
\bar{\psi}_i \gamma^3\gamma^{\mu} \psi^j 
\bar{\psi}_j \gamma^3\gamma^{\mu}\psi^i \nn \\
& &
- \bar{\psi}_i \gamma^5\gamma^{\mu} \psi^j 
\bar{\psi}_j \gamma^5\gamma^{\mu}\psi^i  
+ \bar{\psi}_i \tau \gamma^{\mu}\psi^j 
\bar{\psi}_j \tau \gamma^{\mu}\psi^i \nn \\
 &=& 2 \bar{\chi}_I\gamma^{\mu}\chi^J \bar{\chi}_J\gamma^{\mu}\chi^I 
\label{eq.40}
\eea
These operators are induced by radiative corrections. However
it is found by the Fierz transformation 
that two of them are independent.
We choose ${\cal O}_S$ and ${\cal O}_P$ as the independent ones
and always rewrite others by using the Fierz transformation,
whenever these are induced.

Before going into the ERG analysis of QED$_3$, let us briefly
summarize the results obtained by other methods.
Appelquist {\it et al}\cite{qed3} examined the SD equations for
the chiral symmetry breaking mass in the
ladder approximation and found the novel phase transition.
They claimed that there are two phases depending only on the number
of flavors $N$ and if $N$ is less than the critical value  
$N_c=32/\pi^2 \sim 3.2$, then the chiral symmetry is spontaneously
broken. In any cases there are a single phase chirally broken or
unbroken.
In this analysis they approximated the photon self-eneregy
by it's large N leading part. Also it is assumed that parity is
not broken a la Vafa-Witten's theorem.
After this work many works have been devoted to improvement of the
approximations and the similar results have been obtained\cite{others}.
On the other hand the MC simulation of the noncompact lattice 
QED also has been examined. 
E.Dagotto {\it et al}\cite{lattice} reported the qualitatively same 
results as the above. The critical number was estimated as
$N_c = 3.5 \pm 0.5$.

Now we consider to apply ERG to this system.
We adopt the following scheme of approximation.
First we truncate the set of induced operators and 
restrict the effective lagrangian to 
\be
{\cal L}_{\rm eff} =
\sum_{i=1}^{N} \bar{\psi}_i(\dsla + e \Asla)\psi^i
+ \frac{1}{4} F_{\mu \nu}^2 -
  \frac{1}{2\xi}(\partial_{\mu}A_{\mu})^2
 - \frac{G_S}{2} {\cal O}_S - \frac{G_P}{2} {\cal O}_P.
\label{eq.41}
\ee
Therefore the RG flows are given in the three dimensional
coupling space spanned by $(e^2, G_S, G_P)$.
In the RG approach we may naturally incorporate all the theories
with the same symmetries, 
namely QED$_3$ with four-fermi interactions\cite{4fermiQED},
simultaneously.

Since our purpose is  to see the chiral phase structure of
QED$_3$, we adopt sharp momentum cutoff preserving the 
chiral U(2N) and parity symmetry, at the
cost of the gauge invariance. 
Here we simply discard the gauge non-invariant corrections,
{\it e.g.} photon mass, induced in such regularization scheme.
Of course the gauge invariant scheme\cite{gauge} is preferable 
to see non-perturbative dynamics by gauge interactions.
Here we would postpone the gauge invariant analysis to the future 
studies.
Rather we use one-loop perturbative results for the gauge
beta function and the fermion anomalous dimension 
as the first step of approximation.

Thus we may simply solve the RG equations for the four-fermi
couplings $(G_S, G_P)$ coupled to the gauge beta function,
\be
\frac{d e^2}{dt} = e^2 - \frac{N}{8}e^4,
\label{eq.42}
\ee
where $t=\ln(\Lambda_0/\Lambda)$. 
The first term represents the canonical scaling of the gauge 
coupling with dimension one half. 
It should be noted that there appears an IR stable fixed point 
(FP) at $e^2=e^{2*}=8/N$.
As is seen later on this FP plays an essential
role for the novel phase transition.
Indeed the FP obtained by the perturbative beta function 
may not be reliable for N not large.
Therefore our analysis as well as the SD analyses is not totally 
confidential.
\footnote{
In this point also the ERG applicable for the non-perturbative
RG equations for the gauge couplings\cite{gauge} is strongly 
desired.}
However as long as existence of the fixed point is assumed,
our results are supposed to be qualitatively correct.

The beta functions for the four fermi couplings $G_S$ and $G_P$
are evaluated by summing up the corrections described in Fig.~2.
In each four-fermi vertex of the diagram the opertaor
${\cal O}_S$ or ${\cal O}_P$ is inserted.
It has been found also that the ladder approximation frequently 
used in the SD approach can be reproduced by restricting them to 
the corrections in the first two lines.

The beta functions for the four-fermi couplings in the ladder
approximation turn out to be
\bea
\dot{G}_S &=& -G_S + \frac{1}{\pi^2} 
\left[
G_S^2 - G_S G_P + \frac{1}{3} G_P^2 + 2 e^2 G_S - \frac{4}{3} e^2 G_P
+ \frac{2}{3}e^4
\right], \nn \\
\dot{G}_P &=& -G_P + \frac{1}{\pi^2} 
\left[
-\frac{1}{6}G_P^2 - \frac{2}{3}e^2 G_P - \frac{2}{3}e^4
\right],
\label{eq.43}
\eea
in the landau gauge $\xi=0$.
It is found that the flow equations for $G'_S = G_S - (1/2)G_P$ 
and $G_P$ are completely decoupled. 
Therefore we may solve the coupled equations for $G'_S$:
\begin{figure}[h]
\begin{minipage}{60mm}
\be
\dot{G}'_S = - G'_S + \frac{1}{\pi^2}
\left( G'_S + e^2 \right)^2,
\label{eq.44}
\ee
and Eq.~(\ref{eq.42}).
The flow diagrams in the $(G'_S, e^2)$ 
plane are shown in Fig.~3
and in Fig.~4 for $N=4$ and for $N=2$ respectively.
It is seen that there appears a crtical surface dividing
into two phases for $N=4$, but not for $N=2$.
We may evaluate the critical value of the flavor number as 
$N_{\rm cr} = 32/\pi^2$, where the two phase structure collapses to 
the single phase.
This critical number coincides with the value obatined by solving
the SD equations in the ladder approximation with Landau gauge\cite{3dqed}. 
The two phases appearing for $N \geq N_{\rm cr}$ are supposed
as chiral symmetry broken and (un)broken phases\cite{Kubota}.
The theories in the unbroken phase turns out to be scale invariant
since they are subject to the IR fixed point.
For $N < N_{\rm cr}$ the fixed point disappears and there 
remains only the broken phase.
\end{minipage}
\hspace*{5mm}
\begin{minipage}{60mm}
\begin{center}
\epsfxsize=1.0\textwidth
\leavevmode
\epsffile{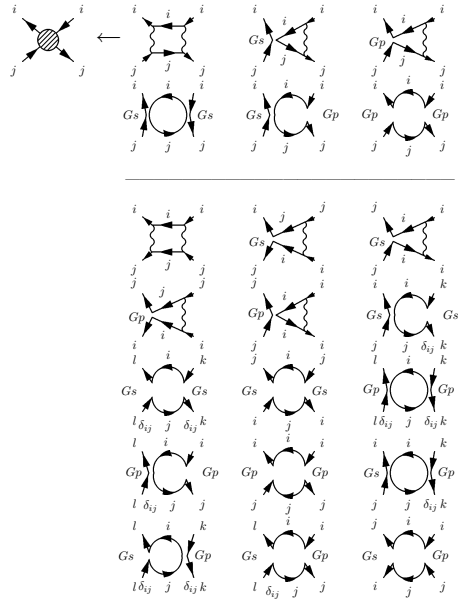}
\caption{The corrections for the four-fermi operators.
The arrows stand for contraction of the spinor indices.}
\end{center}
\end{minipage}
\end{figure}

\begin{figure}[h]
\begin{center}
\begin{minipage}[t]{55mm}
\epsfxsize=1.0\textwidth
\leavevmode
\epsffile{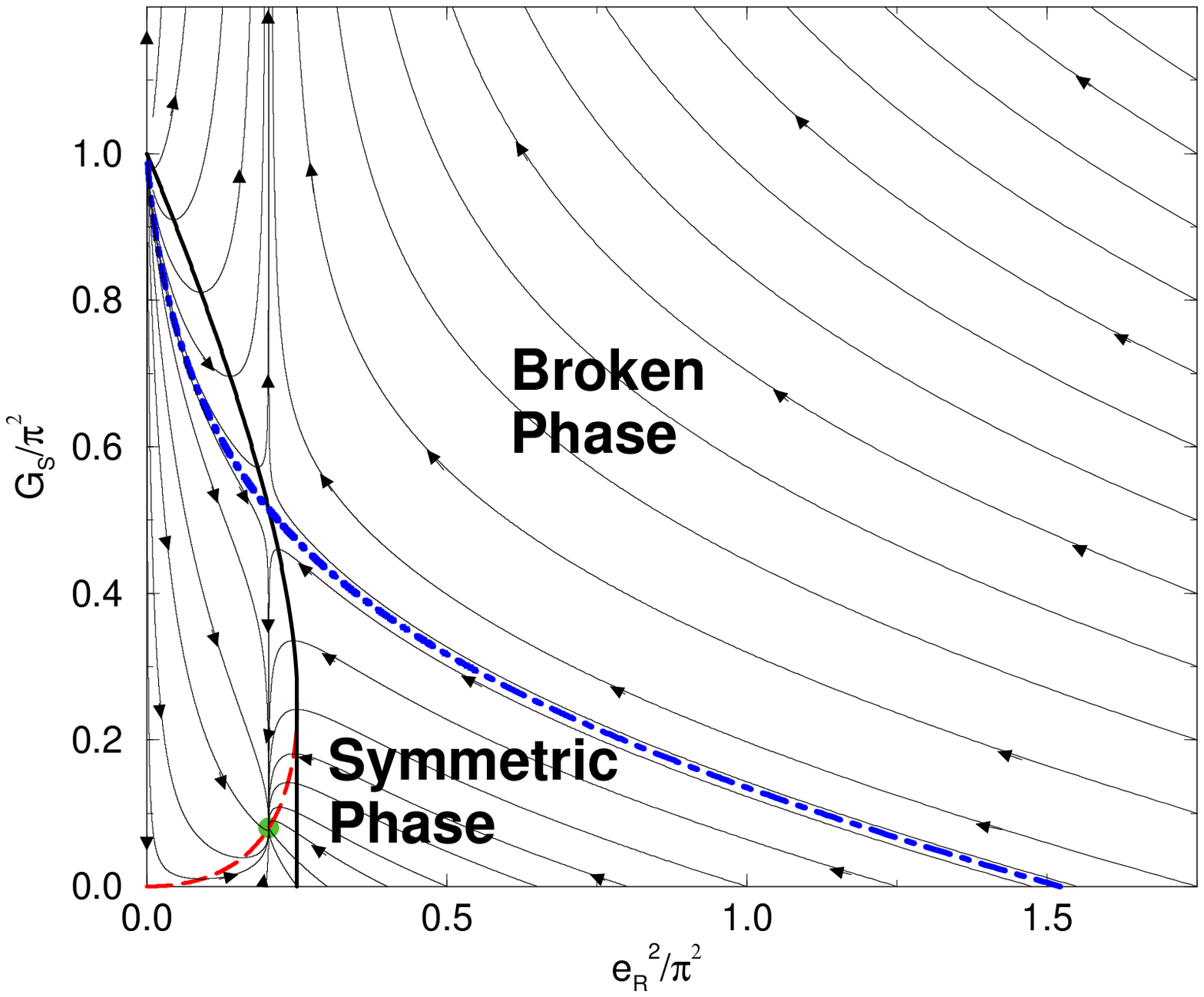}
\caption{RG flows of $(G'_S, e^2)$ for N=4.}
\end{minipage}
\leavevmode
\hspace*{10mm}
\begin{minipage}[t]{55mm}
\epsfxsize=1.0\textwidth
\epsffile{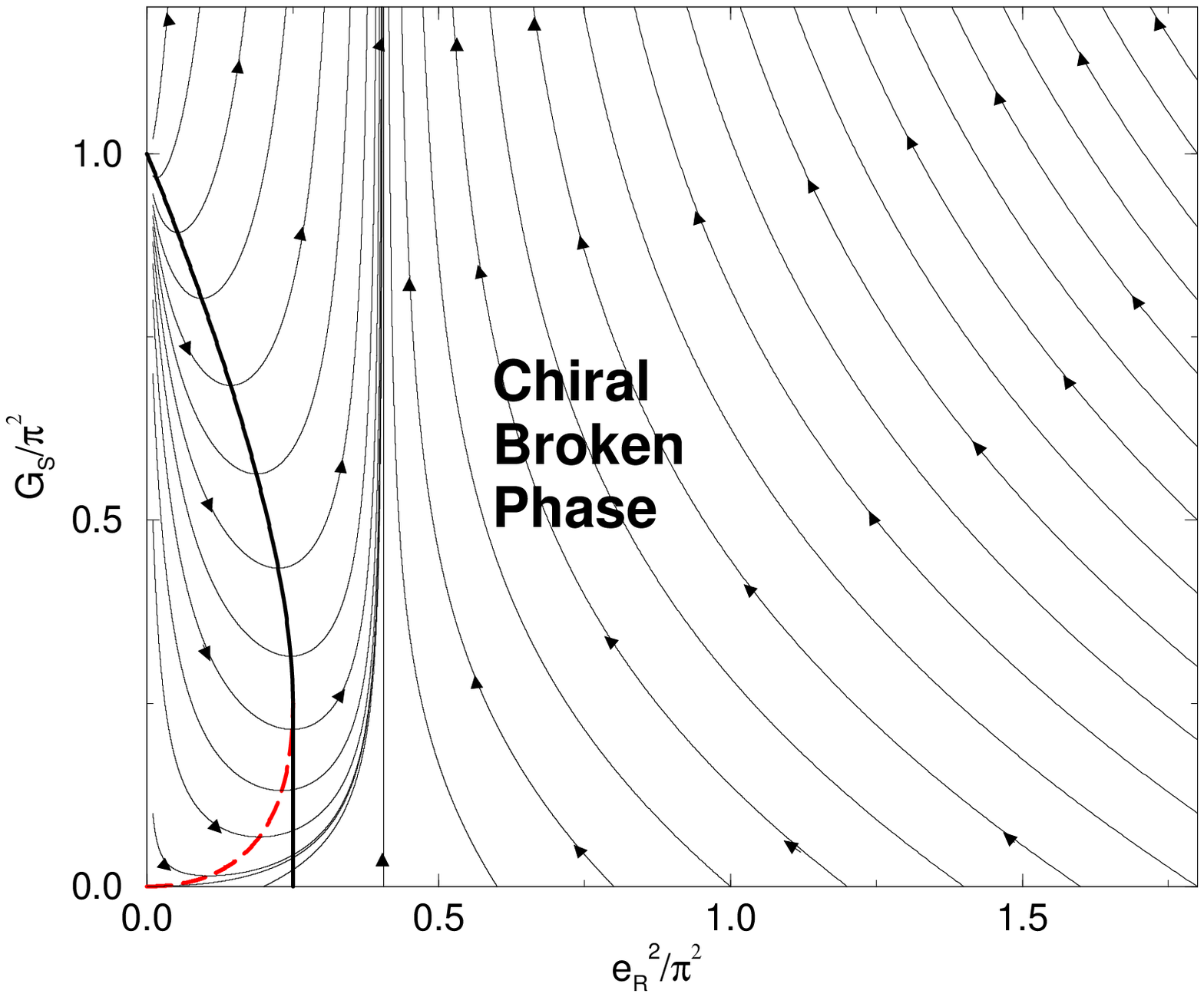}
\caption{RG flows of $(G'_S, e^2)$ for N=2.}
\end{minipage}
\end{center}
\end{figure}

Here we should note that our results are slightly different from
them claimed by the MC simulation and by the SD analyses,
since they tell that no symmetry breaking occurs for
$N \geq N_{\rm cr}$.
Actually the continuum limit of the
theory is studied in these analyses. 
In the RG point of view we may take the
continuum limit only when the gauge coupling is less than
the IR fixed point value\cite{Kubota}. 
If we restrict to such cases, the theories always lie in 
the unbroken phase for $N \geq N_{\rm cr}$.
Thus our results are not conflicting with them.
However we may well regards QED$_3$ as an effective theory with
a certain underlying cutoff. 
Then the chiral symmetry is always
spontaneously broken when the gauge coupling is strong enough.

It is a great advantage of the ERG approach to enable us to
incorporate all the corrections shown in Fig.~2 quite
easily. This benefit is not solely the matter of improvement 
from the ladder approximation, but also saving from the large
gauge dependence.
The reason is similar with that the gauge independence of the 
on-shell S-matrix is achieved by summing up all the diagram 
appearing in each loop order. 
In our RG equations the one-loop corrections to the four-fermi
interactions are completed diagramatically by adding the
corrections to the external fermion legs to the full set of
the diagrams shown in Fig.~2.
This effect may be incorporated by taking the anomalous 
dimension of fermions into account.
If we evaluate the anomalous dimension also by one-loop 
perturbation, then we may obtain
the gauge independent beta functions, 
which are found out to be
\bea
\dot{G}_S &=& -G_S + \frac{1}{\pi^2} 
\left[
\frac{N+2}{3}G_S^2 - G_S G_P + \frac{4}{3} e^2 G_S - \frac{8}{3} e^2 G_P
\right], \nn \\
\dot{G}_P &=& -G_P + \frac{1}{\pi^2} 
\left[
(2N-1)G_P^2 -2(N-1)G_S G_P- \frac{4N-7}{6}G_S^2 \right. \nn \\
& &\left.  - \frac{8}{3}e^2 G_S + \frac{4}{3}e^2 G_S - 2 e^4
\right].
\label{eq.45}
\eea

Numerical analysis tells us that there remains the critical 
number of flavors at $3 < N_{\rm cr}< 4$.
In Fig.~5 and Fig.~6 the RG flows 
of $(G_S, G_P)$ couplings on the plane of $e^2=e^{2*}$ are shown for $N=4$
and $N=2$ respectively.
\begin{figure}[h]
\begin{center}
\begin{minipage}[t]{55mm}
\epsfxsize=1.0\textwidth
\leavevmode
\epsffile{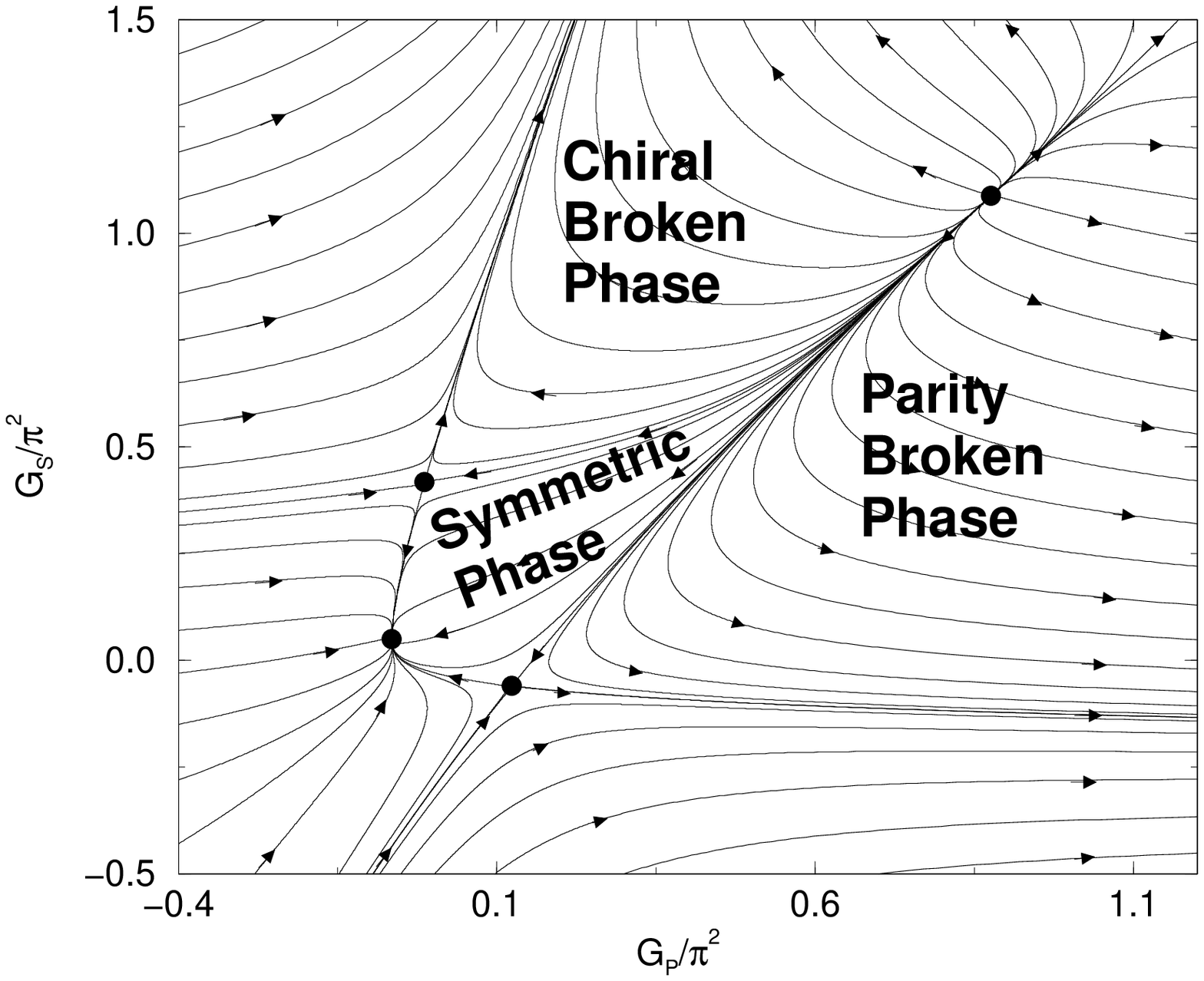}
\caption{RG flows of $(G_S, G_P)$ on the plane of 
$e^2=e^{2*}$ are shown for $N=4$.}
\end{minipage}
\hspace*{10mm}
\leavevmode
\begin{minipage}[t]{56mm}
\epsfxsize=1.0\textwidth
\epsffile{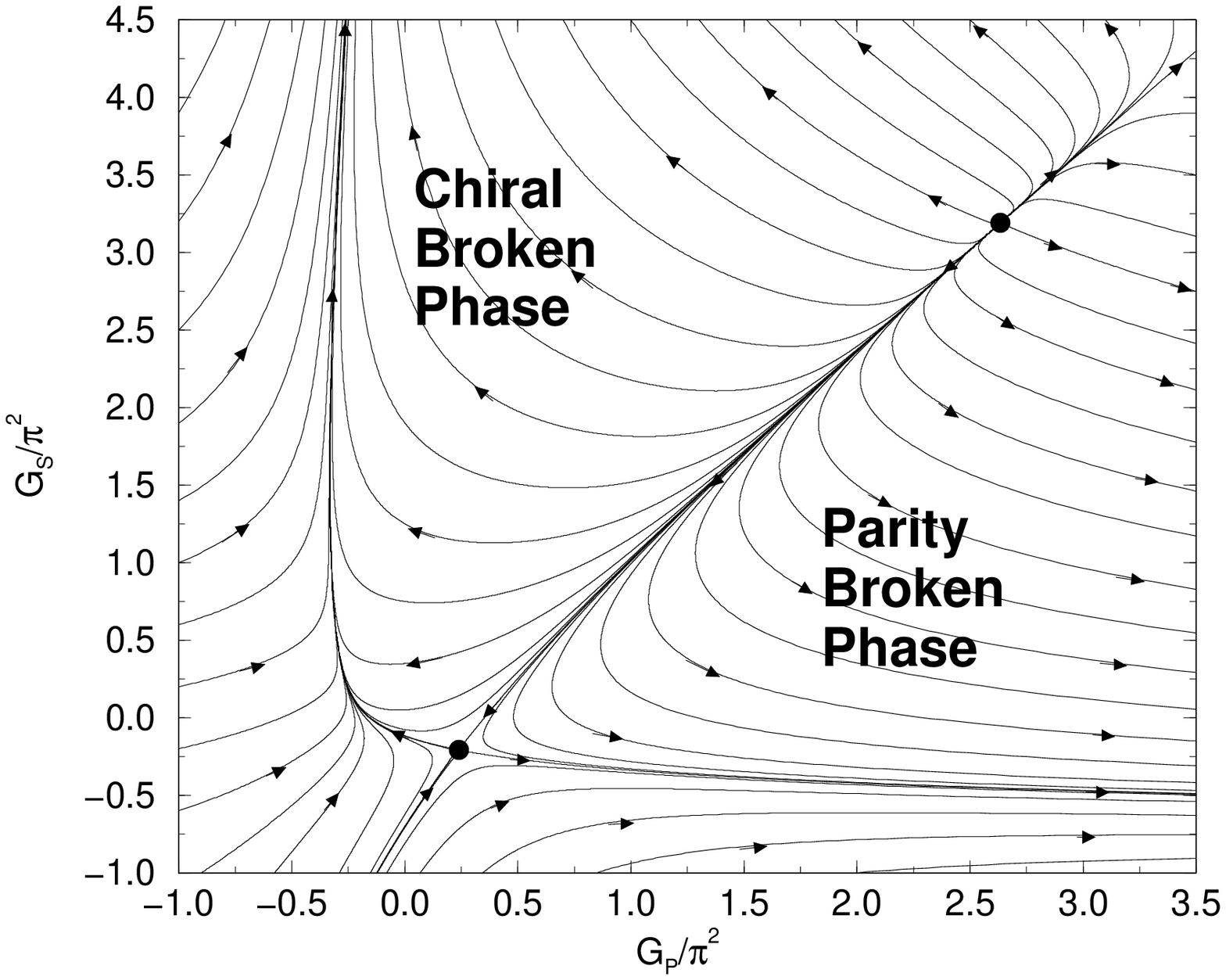}
\caption{RG flows of $(G_S, G_P)$ on the plane of 
$e^2=e^{2*}$ are shown for $N=2$.}
\end{minipage}
\end{center}
\end{figure}
It is seen that there are three phases, which are supposed to be
symmetric, chiral symmetry broken and parity broken phases, for
N larger than the critical value.
As N becomes smaller than the critical value the symmetric phase
disappears.
It is also seen that the RG flows of QED$_3$ always run outside
of the parity broken phase, which is consistent with Vafa-Witten's
theorem.
However parity can be broken for QED$_3$ with the general 
four-fermi interactions. 
Besides it is seen that the tricritical fixed point appears 
at the edge of the boundary between the chiral broken phase and the
parity broken phase.
This implies that the phase transition turns to first order beyond
this edge.
Thus we are able to grapse the phase structure of dynamical
symmetry breakings easily by means of ERG. 
This is a remarkable point of ERG, though we cannot assert 
which symmetry is broken in each phase only from the RG flows 
of the four-fermi couplings. 
Therefore we may well expect that ERG method will be effective
also to other interesting cases, {\it e.g.} color
superconductivity.

\section{Conclusions}
First we considered mainly the formal aspects of 
the ERG in comparison with the SD equations.
Their explicit relations are given in large N vector models
and in the Gross-Neveu model.
It has been shown formally that the ERG equations and the SD
equations give the identical ganerating functional as their 
solution.
It was also shown that the ERG method is applicable to the
large N matrix models, where radiative corrections given by 
the planar diagrams are easily taken by solving the approximated
RG equations.

In the second part, we analyzed the phase structure of QED$_3$
by applying the ERG in the primitive level of approximations.
By considering the RG flow equations for the four-fermi
interactions, we could understand the novel phase transition
advocated by the SD approach and also by the lattice simulations
after quite simple calculations. 
The resultant RG flows show the phase structure of chiral and parity
symmetries immediately.
It is also noted that we do not need to assume the symmetry to be 
broken apriori contrary to the SD approach.
These observations demonstrate that ERG predominates over the
SD methods in clarifying complicated phase diagrams of
dynamical symmetry breaking.
It is indeed true that our analysis is not confidential due to
poor treatment of the RG equation for the gauge coupling.
We would like to expect future development in this direction.

\section*{Acknowledgments}
The author would like to
thank the organizers of the sitimulating conference, 
S.~Arnone, T.~R.~Morris and K.~Yoshida.
He is also indebted to K.-I.~Aoki, K.-I.~ Kubota,
K.~Morikawa, H.~Onoda, J.-I.~ Sumi, K.~Takagi, 
M.~Tomoyose for fruitful collaboration.

\end{document}